\documentclass[preprint]{aastex631}

\usepackage[T1]{fontenc}

\newcommand{\sfrac}[2]{\mathchoice%
  {\kern0em\raise.5ex\hbox{\the\scriptfont0 #1}\kern-.15em/
    \kern-.15em\lower.25ex\hbox{\the\scriptfont0 #2}}
  {\kern0em\raise.5ex\hbox{\the\scriptfont0 #1}\kern-.15em/
    \kern-.15em\lower.25ex\hbox{\the\scriptfont0 #2}}
  {\kern0em\raise.5ex\hbox{\the\scriptscriptfont0 #1}\kern-.2em/
    \kern-.15em\lower.25ex\hbox{\the\scriptscriptfont0 #2}} {#1\!/#2}}

\usepackage{amsmath}

\newcommand{\C}{$^{12}$C}

\usepackage{bm}

\begin{document}

\title{Well-Balanced Hydrodynamics for the Piecewise Parabolic Method with Characteristic Tracing}

\shorttitle{Well-balanced PPM Hydrodynamics}

\author[0000-0001-8401-030X]{Michael Zingale}
\affiliation{Department of Physics and Astronomy, Stony Brook University,
             Stony Brook, NY 11794-3800, USA}

\correspondingauthor{Michael Zingale}
\email{michael.zingale@stonybrook.edu}

\begin{abstract}
Well-balanced reconstruction techniques have been developed for stellar hydrodynamics
to address the challenges of maintaining hydrostatic equilibrium during evolution.
{I} show how to adapt a simple well-balanced method to the piecewise
parabolic method for hydrodynamics.  A python implementation of the method
is provided.
\end{abstract}


\section{Introduction}\label{Sec:Introduction}

The piecewise parabolic method (PPM) \citep{ppm} is widely used in
astrophysical hydrodynamic simulation codes due to its low numerical
viscosity.  PPM reconstructs data on a grid as parabolas and
integrates under them to accumulate the information that can make it
to an interface over a timestep---this procedure is called
characteristic tracing.  This is then used to
compute the fluxes through the zones and update the state in time.  An
alternative to characteristic tracing is a method-of-lines
discretization, where the parabolic reconstruction is only used to
find the value on the interface, and the time integration is treated
as a system of ordinary differential equations.

A common problem with stellar hydrodynamics is maintaining hydrostatic
equilibrium (HSE)---this requires the exact cancellation of the pressure
gradient and the gravitational source, otherwise an acceleration will
be generated.  Well-balanced method{s} take the hydrostatic profile into
account when doing the reconstruction.  Recent techniques
\citep{kappeli:2016, kappeli:2022} have shown how to achieve HSE to machine
precision, but these focus piecewise linear reconstruction and method-of-lines integration.  Here
{I} demonstrate how to apply the same ideas to the original characteristic
tracing formulation of PPM.

\section{Well-Balanced PPM}

PPM reconstruction works with the primitive
variable formulation of the Euler equations:
\begin{equation}
    {\bf q}_t + {\bf A}({\bf q}) {\bf q}_x + {\bf H}
\end{equation}
where ${\bf q} = (\rho, u, p)^\intercal$ (density, velocity, and
pressure), ${\bf H} = (0, g, 0)^\intercal$ is the gravitational source
term, ${\bf A}$ is
\begin{equation}
    {\bf A} = \left ( \begin{array}{ccc} u & \rho & 0 \\
                                         0 & u & 1/\rho \\
                                         0 & \Gamma_1 p & u \end{array} \right )
\end{equation}
and $\Gamma_1$ is the adiabatic index.
The goal is to predict the values of ${\bf q}$ on each side of each
interface separating the zones and then compute the flux through the
zones by solving a Riemann problem.  This interface state prediction,
described in \citet{ppm}, proceeds as:
\begin{enumerate}
\item Use a conservative cubic interpolant to find the interface
  values ${\bf q}$.  {I'll} write this as ${\bf q}_{i+1/2} = f({\bf
    q}_{i-1}, {\bf q}_{i}, {\bf q}_{i+1}, {\bf q}_{i+2})$

\item For zone $i$, define the left and right edges of a
  parabolic reconstruction, ${\bf q}_{-,i}$ and ${\bf q}_{+,i}$:
 \begin{equation}
     {\bf q}_{-,i} = {\bf q}_{i-1/2} \, ;\quad  {\bf q}_{+,i} = {\bf q}_{i+1/2}
 \end{equation}
 These values are then limited according to the procedure in
 \citet{ppm}, resulting
 in a parabola for zone $i$ that {I'll} denote ${\bf q}(x)$.

\item For each of the characteristic waves, $\lambda^{(\nu)} = \{u -
  c, u, u + c\}$ ({I'll} label these as $-$, $0$, and $+$, respectively), define the integral under the parabola from each edge, covering the
  domain the wave sees in a timestep $\Delta t$:
  \begin{equation}
  \mathcal{I}_+^{(\nu)}({\bf q}_i) =
      \frac{1}{\sigma_i^{(\nu)} \Delta x} \int_{x_{i+1/2} - \sigma_i^{(\nu)}\Delta x}^{x_{i+1/2}} {\bf q}(x) \, dx
  \end{equation}
and
  \begin{equation}
  \mathcal{I}_-^{(\nu)}({\bf q}_i) =
      \frac{1}{\sigma_i^{(\nu)} \Delta x} \int_{x_{i-1/2}}^{x_{i-1/2} + \sigma_i^{(\nu)}\Delta x} {\bf q}(x) \, dx
  \end{equation}
where $\sigma_i^{(\nu)}$ is the effective Courant number of characteristic wave $\nu$, $\sigma_i^{(\nu)} = {\lambda_i^{(\nu)}\Delta t} / {\Delta x}$.

\item The final states on each interface seen by zone $i$ are
  constructed by projecting the jumps carried by each wave into characteristic variables using the
  left and right eigenvectors of ${\bf A}$, ${\bf l}^{\nu}$ and ${\bf
    r}^{\nu}$, and summing these contributions.  This is done with respect to a reference
    state---$\mathcal{I}({\bf q})$ of the fastest wave moving toward the interface:
\begin{align}
  {\bf q}_{i+1/2,L}^{n+1/2} &= \mathcal{I}_+^{(+)}({\bf q}_i) -
   \sum_{\nu;\lambda^{(\nu)}\ge 0} {\bf l}_i^{(\nu)} \cdot \left (
        \mathcal{I}_+^{(+)}({\bf q}_i) - \mathcal{I}_+^{(\nu)}({\bf q}_i) -
                            \frac{\Delta t}{2} \mathcal{I}_+^{(\nu)}({\bf H}_i)
       \right ) {\bf r}_i^{(\nu)} \\
  {\bf q}_{i-1/2,R}^{n+1/2} &= \mathcal{I}_-^{(-)}({\bf q}_i)-
   \sum_{\nu;\lambda^{(\nu)}\le 0} {\bf l}_i^{(\nu)} \cdot \left (
        \mathcal{I}_-^{(-)}({\bf q}_i) - \mathcal{I}_-^{(\nu)}({\bf q}_i) -
                            \frac{\Delta t}{2} \mathcal{I}_-^{(\nu)}({\bf H}_i)
       \right ) {\bf r}_i^{(\nu)}
\end{align}
Note that this prescription includes the effects of the source, ${\bf H}$ over $\Delta t$.
\end{enumerate}

The final interface state is found by solving the Riemann problem,
\begin{equation}
{\bf q}_{i+1/2}^{n+1/2} = \mathcal{R}({\bf q}_{i+1/2,L}^{n+1/2}, {\bf q}_{i+1/2,R}^{n+1/2})
\end{equation}
and in the conservative update, HSE will appear as:
\begin{equation}
\frac{p_{i+1/2}^{n+1/2} - p_{i-1/2}^{n+1/2}}{\Delta x} = \frac{1}{2} (\rho_i^n + \rho_i^{n+1}) g \enskip .
\end{equation}
In the standard PPM method, these do not cancel for an HSE atmosphere due to truncation error.

\cite{kappeli:2016} showed that {one can} subtract the reconstructed HSE pressure
from the pressure before doing slope limiting in a piecewise linear method.  As
long as the HSE reconstruction method matches the discretization with which the initial model was prepared, this will preserve HSE to roundoff level during evolution.
That prescription worked with method-of-lines integration.  Here {I} show how
to use the same ideas with PPM and characteristic tracing.  Our approach differs
from the PPM reconstruction in \citet{ppm-hse}---there $(\rho g)$ was reconstructed
in each zone as a parabola and used to modify the pressure.

Working on zone $i$ and building the interface states it influences,
the modifications to the PPM reconstruction are:
\begin{itemize}
    \item Define the perturbational pressure, $p'$, in the surrounding
      zones, by subtracting off the hydrostatic pressure:
    \begin{align}
       p'_i &= 0 \\
       p'_{i\pm 1} &= p_{i\pm 1} - \int_{x_i}^{x_{i\pm 1}} \rho g dx\\
       p'_{i\pm 2} &= p_{i\pm 2} - \int_{x_i}^{x_{i\pm 2}} \rho g dx
    \end{align}
    where, following \citet{kappeli:2016}, {I} use a piecewise constant profile for $\rho$ and $g$ in each
    zone.
    \item Do the parabolic reconstruction on $p'$.  For zone $i$, this means that {the initial left and right interface states need to be computed} with the same HSE state:
     \begin{align}
         p'_{-,i} = p'_{i-1/2} &= f(p'_{i-2}, p'_{i-1}, p'_i, p'_{i+1}) \\
         p'_{+,i} = p'_{i+1/2} &= f(p'_{i-1}, p'_{i}, p'_{i+1}, p'_{i+2}) 
     \end{align}
     and then use these to define the parabola, $p'(x)$, and define the hydrostatic
     pressure on edges:
     \begin{align}
         p^\mathrm{HSE}_{-,i} &= p_i - \frac{\Delta x}{2} \rho_i g_i \\
         p^\mathrm{HSE}_{+,i} &= p_i + \frac{\Delta x}{2} \rho_i g_i
     \end{align}

    \item Perform the characteristic tracing with $p'$ and do not
      include the velocity source terms in the tracing (consistent with \citealt{ppm-hse}).  After the tracing, the
      hydrostatic pressure needs to be added to the traced
      perturbational pressure.  This gives:
      \renewcommand{\arraystretch}{1.25}
       \begin{align}
  \left ( \begin{array}{c} \rho_{i+1/2,L}^{n+1/2} \\
                           u_{i+1/2,L}^{n+1/2} \\
                           p_{i+1/2,L}^{n+1/2} \end{array} \right )
  &= \left ( \begin{array}{c} \mathcal{I}_+^{(+)}(\rho_i) \\
                              \mathcal{I}_+^{(+)}(u_i) \\
                              p^\mathrm{HSE}_{+,i} + \mathcal{I}_+^{(+)}(p'_i) \\
                    \end{array} \right ) -
   \sum_{\nu;\lambda^{(\nu)}\ge 0} {\bf l}_i^{(\nu)} \cdot
        \left ( \begin{array}{c} \mathcal{I}_+^{(+)}({\rho}_i) - \mathcal{I}_+^{(\nu)}({\rho}_i) \\
        \mathcal{I}_+^{(+)}({u}_i) - \mathcal{I}_+^{(\nu)}({u}_i)\\
        \mathcal{I}_+^{(+)}(p'_i) - \mathcal{I}_+^{(\nu)}(p'_i) \end{array} \right )
        {\bf r}_i^{(\nu)} \\
  \left ( \begin{array}{c} \rho_{i-1/2,R}^{n+1/2} \\
                           u_{i-1/2,R}^{n+1/2} \\
                           p_{i-1/2,R}^{n+1/2} \end{array} \right )
  &= \left ( \begin{array}{c} \mathcal{I}_-^{(-)}(\rho_i) \\
                              \mathcal{I}_-^{(-)}(u_i) \\
                              p^\mathrm{HSE}_{-,i} + \mathcal{I}_-^{(-)}(p'_i) \\
                    \end{array} \right ) -
   \sum_{\nu;\lambda^{(\nu)}\le 0} {\bf l}_i^{(\nu)} \cdot
        \left ( \begin{array}{c} \mathcal{I}_-^{(-)}({\rho}_i) - \mathcal{I}_-^{(\nu)}({\rho}_i) \\
        \mathcal{I}_-^{(-)}({u}_i) - \mathcal{I}_-^{(\nu)}({u}_i)\\
        \mathcal{I}_-^{(-)}(p'_i) - \mathcal{I}_-^{(\nu)}(p'_i) \end{array} \right )
        {\bf r}_i^{(\nu)}
\end{align}
\renewcommand{\arraystretch}{1.0}

    \citet{ppm} evaluate ${\bf l}^{(\nu)}$ and ${\bf r}^{(\nu)}$ using the reference states,
    so the HSE pressure would need to be added back first for that evaluation.
\end{itemize}

{I} provide an implementation of this method in a simple 1D PPM
hydrodynamics code, {\sf
  PPMpy}\footnote{https://github.com/python-hydro/ppmpy} \citep{ppmpy}.  
This method is also being made available in {\sf Castro} \citep{castro_joss}.
\begin{figure}
    \plottwo{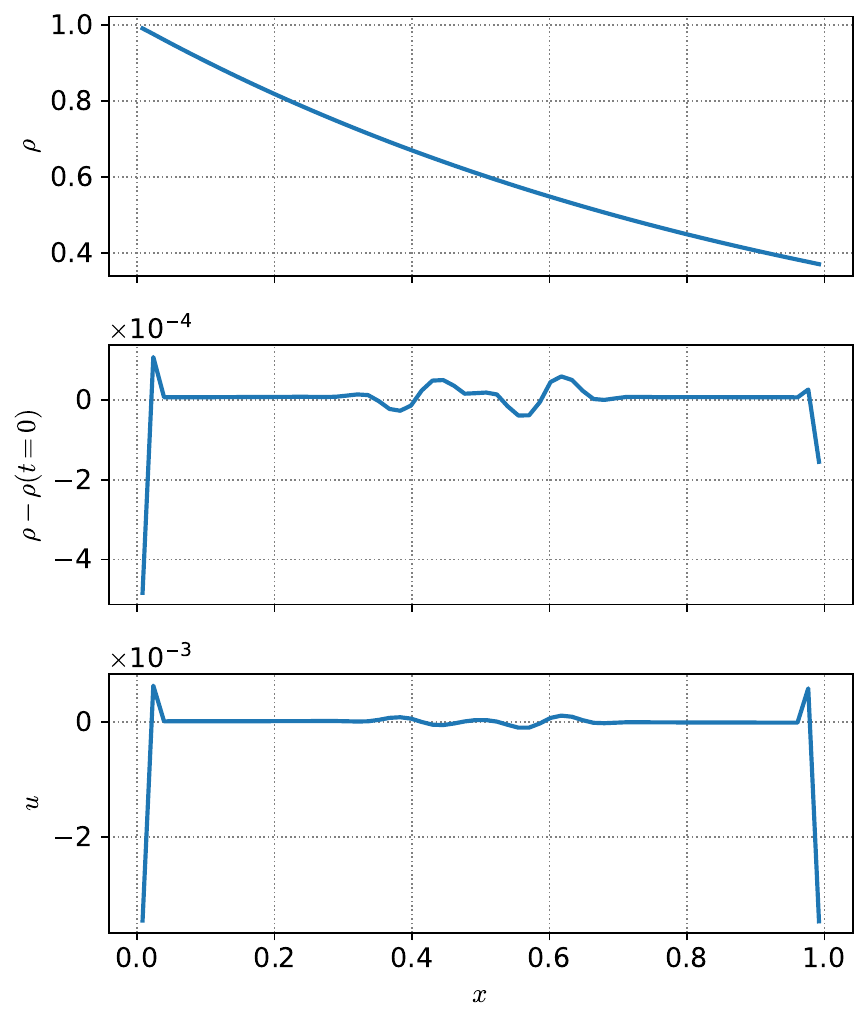}{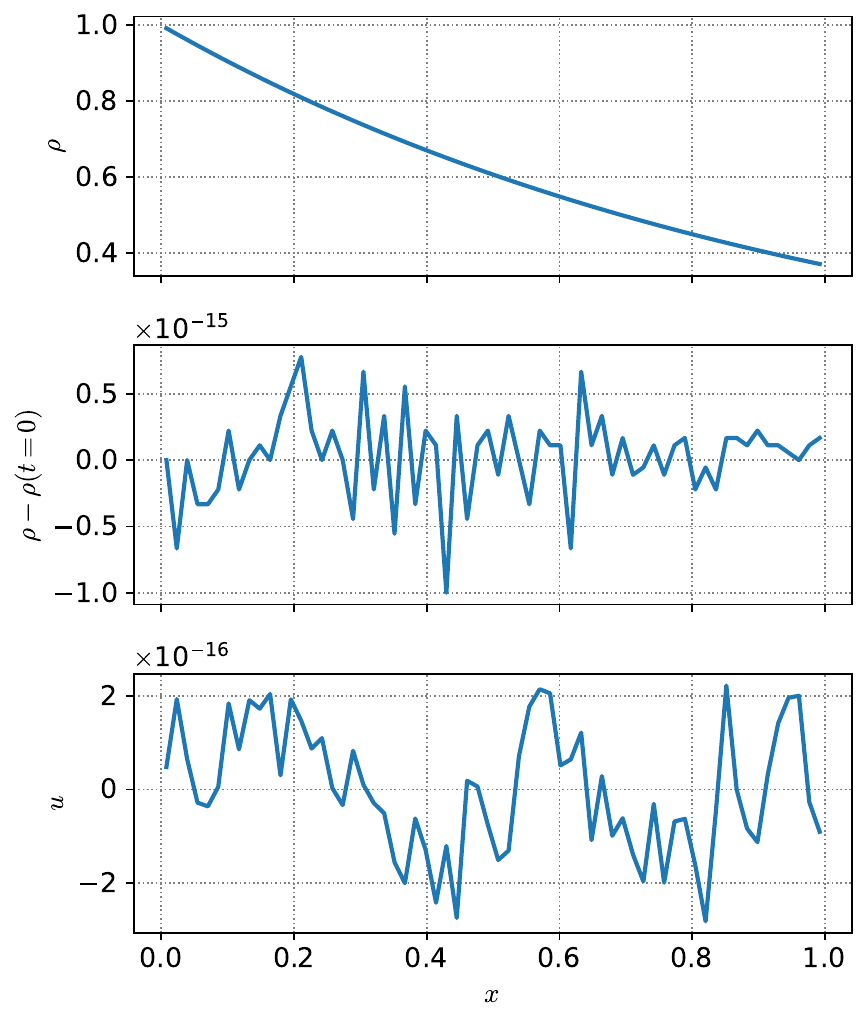}
\caption{\label{fig:hse} Density, change in density over evolution, and final velocity
  for the standard PPM reconstruction (left) and this well-balanced method (right).}
\end{figure}

To test the method, {I} explore an isothermal hydrostatic
atmosphere, with base density of $1$, pressure of $1$, and $g = -1$ in a domain
$[0, 1]$ (in code units).  The initial atmosphere is constructed by first
integrating from the lower boundary to the first cell center, and then
differencing HSE as:
\begin{equation}
   p_{i+1} = p_i + \frac{1}{2} \Delta x (\rho_i + \rho_{i+1}) g
\end{equation}
with an isothermal ideal gas.  {This is run} for a time of 0.5 using {\sf PPMpy}{,
with} reflecting boundaries, which work well when the reconstruction is done with the
perturbational pressure.  Figure~\ref{fig:hse} shows the
atmosphere with both the standard PPM reconstruction and this
well-balanced method{ and demonstrates} that with the well-balanced approach,
the velocity is roundoff-level throughout the atmosphere.

\ \\
The work at Stony Brook was supported by DOE/Office of Nuclear
Physics grant DE-FG02-87ER40317. 

\software{matplotlib \citep{Hunter:2007}, numpy \citep{numpy2}}


\bibliographystyle{aasjournal}
\bibliography{ws}

\end{document}